# Free Recall Shows Similar Reactivation Behavior as Recognition & Cued Recall

Eugen Tarnow, Ph.D.

18-11 Radburn Road

Fair Lawn, NJ 07410

etarnow@avabiz.com (e-mail)

16462290787 (phone)

1208 445 3638 (fax)

## Abstract:

I find that the total retrieval time in word free recall increases linearly with the total number of items recalled.  Measured slopes, the time to retrieve an additional item, vary from 1.4-4.5 seconds per item depending upon presentation rate, subject age and whether there is a delay after list presentation or not.  These times to retrieve an additional item obey a second linear relationship as a function of the recall probability averaged over the experiment, explicitly independent of subject age, presentation rate and whether there is a delay after the list presentation or not.  This second linear relationship mimics the relationships in recognition and cued recall (Tarnow, 2008) which suggests that free recall retrieval uses the same reactivation mechanism as recognition or cued recall.  Extrapolation limits the time to retrieve an additional item to a maximum of 7.2 seconds per item if the probability of recall is near 0%. Earlier upper limits for recognition and cued recall varied with the item type between 0.2 and 1.8 seconds per item (Tarnow, 2008). Implications include that there are only two types of short term memory: working memory and reactivation.

Keywords: Free recall; short term memory; memory search; retrieval

## Introduction

Free recall experiments typically display a list of words and then ask the subjects to recall as many of the words as possible. In a very colorful article, Hintzman (2011) wrote of free recall: "the overlay of study and retrieval strategies makes the task a grotesque, neither-fish-nor-fowl creature of the laboratory—corresponding to nothing people do in everyday life and too complex to be of much use." As a result, he noted that interest in free recall peaked in 1970 and has since been low. Recognition, on the other hand, he points out is still of interest to the research community.

In this contribution I will show that free recall, cued recall and recognition are all very similar. This would probably have been discovered earlier had we not started out with the difficult serial-position curve with primacy and recency effects (Fig. 1).

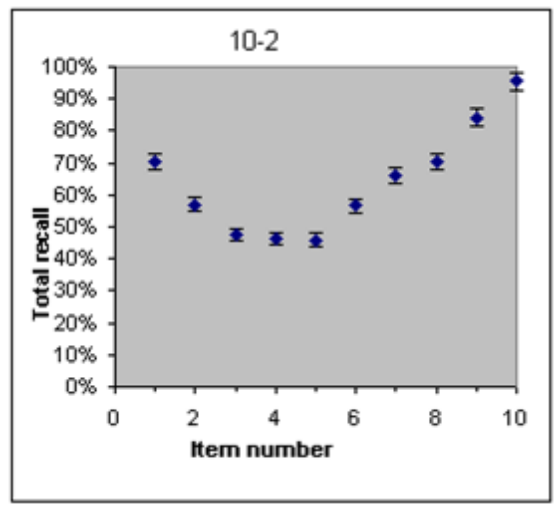

*Fig. 1. The famous bowed curve of total recall versus word number (Murdock, 1962). Ten items were displayed at a rate of one item per two seconds.*

## Results

In the word free recall data of Murdock & Okada (1970) and Kahana, Zaromb & Wingfield (2001) (Fig. 2) I find that there is a linear relationship between the total free recall retrieval time, defined as the sum of all response times up to the last item N, and the total N number of items recalled. The linear fits account for 88-99% of the variance in the word free recall response times. The slopes, the average retrieval time for an additional item, vary between 1.4-4.5 seconds per item and depend upon presentation rate, subject age and whether there is a delay. The times to retrieve additional items are smaller for the slow presentation rate, for young subjects and for presentation without delay. In Fig. 3 is displayed the times to retrieve additional items as a function of the overall probability of recall. They form a second linear relationship. Extrapolation to 0% recall probability gives a longest retrieval time of 7.2 seconds per item. Strangely, a shortest retrieval time of 0 seconds is extrapolated to occur for a 66% recall probability.

I note that one free recall experiment was excluded, the immediate and delayed free recall results from Howard & Kahana (1999). The retrieval times were linear with the total number items recalled (slopes 4.16 and 2.59 seconds for the immediate and delayed free recall) but the trend was the opposite expected: the slope was larger for the immediate free recall than for the delayed free recall. Since this trend was different in Kahana, Zaromb & Wingfield (2001) even though the experimental procedure seemed very similar, I suggest that there is something incorrect with the time registrations of the Howard & Kahana (1999) experiment (repeated emails to verify the data with the authors were not answered).

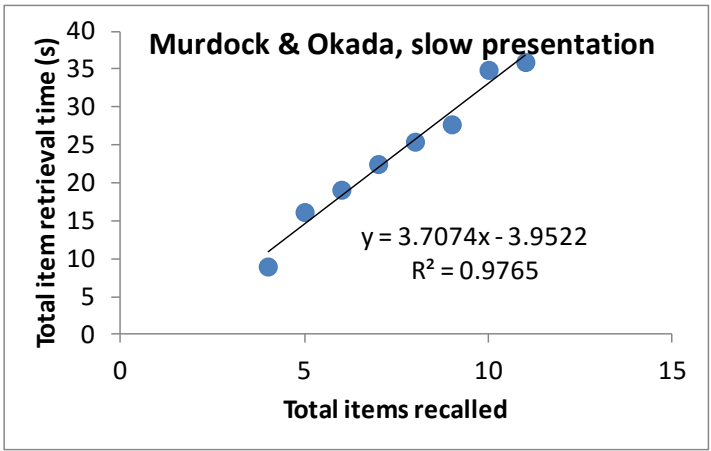
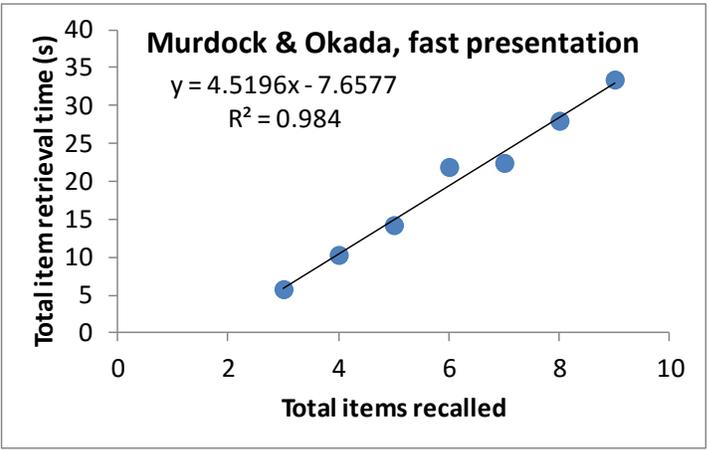
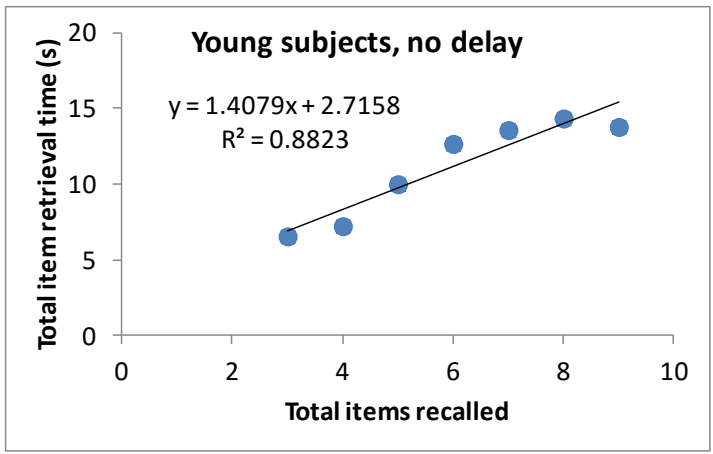
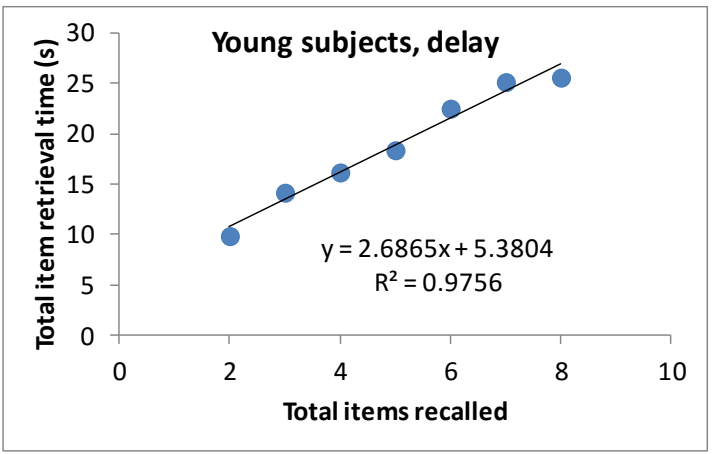
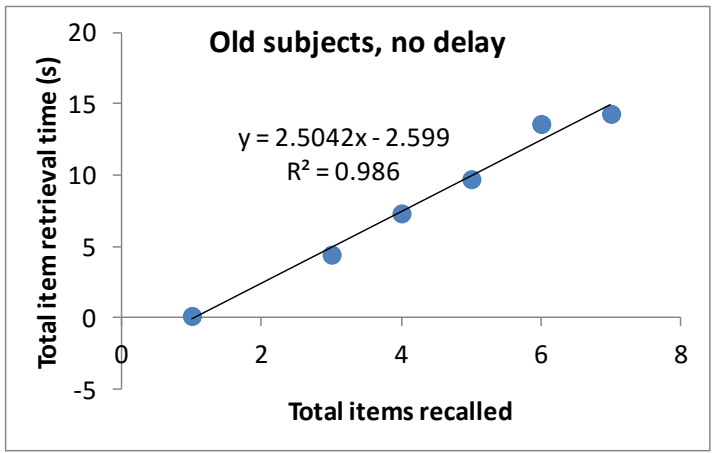
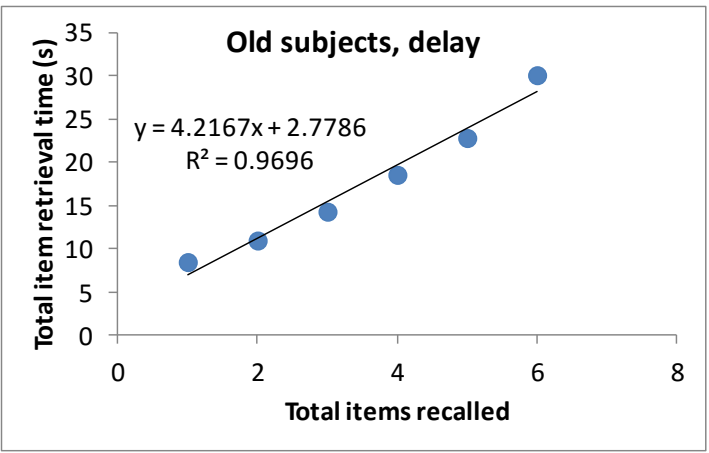

Fig 2. The sum of all retrieval times is a linear function of the total number of items retrieved in word free recall. The top row corresponds to the immediate free recall data for slow (left panel) and fast (right panel) presentations in the Mudock & Okada (1970), the middle and bottom rows to the data in Kahana, Zaromb and Wingfield (2001). To minimize statistical

*noise, points with fewer than 10 values were not included in the graphs above. To make the measurements well defined, no recall series containing errors were included.*

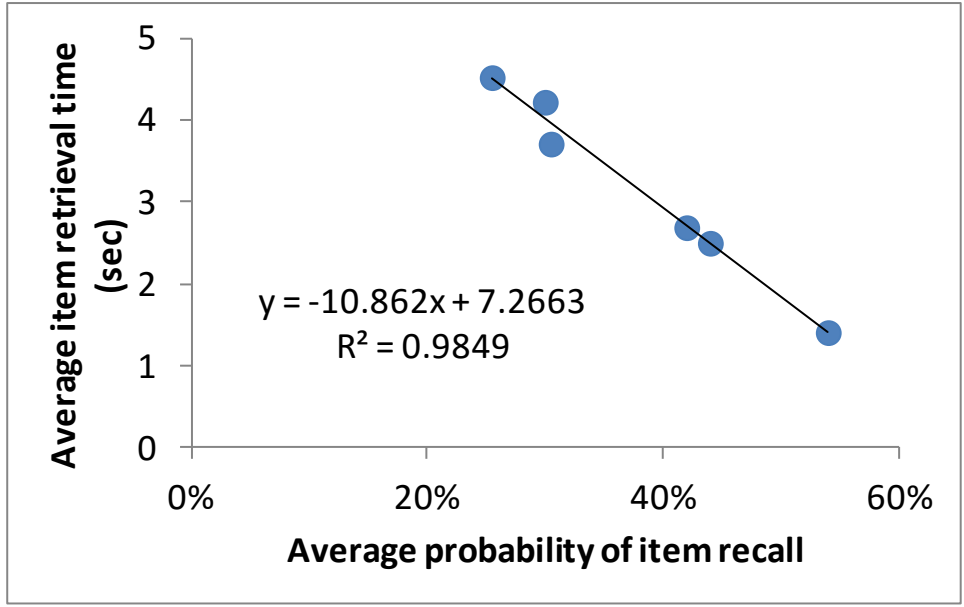

Fig 3. Average time to find an additional item as a function of the average probability of recall using the slopes of Fig. 2. Extrapolation to 0% probability of recall suggests that the largest average time to retrieve an additional item would be 7.2 seconds and 0 seconds if the average probability of item retrieval is 66% or higher.

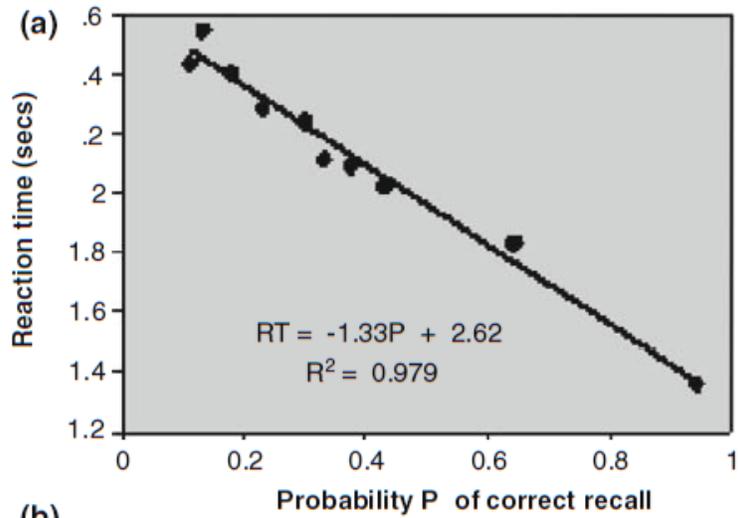

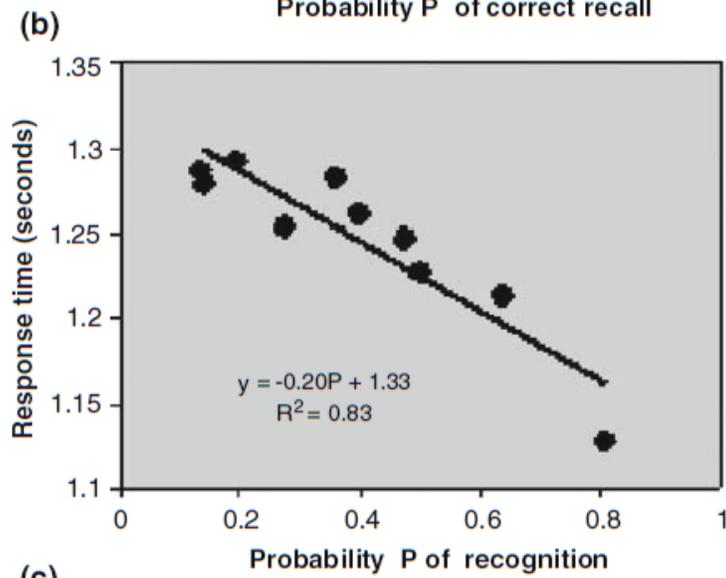

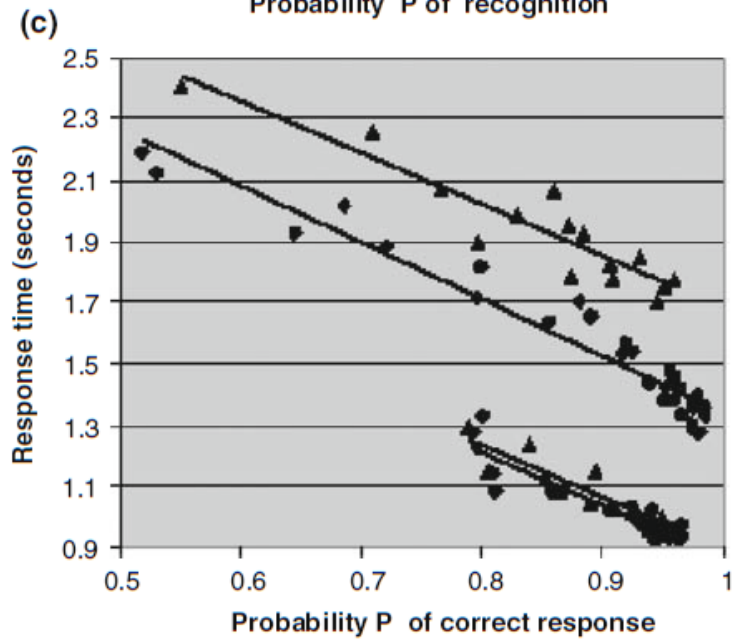

*Fig 4. Figure taken from Fig. 1 of Tarnow (2008). Response times as a function of the probability of recognition / cued recall. Top panel displays cued recall, middle panel displays recognition and bottom panel displays a different dataset with the upper curves from cued recall and the lower curves from recognition.*

## Discussion

I found that the average time to retrieve an additional item in a free recall experiment is a constant.

The linear relationships in Fig. 2 and Fig. 3 are surprising considering that individual item response times vary widely from half a second to ten seconds or even longer.

Importantly, these linear relationships are similar to those I found earlier describing recognition and cued recall (Tarnow, 2008; reproduced in Fig. 4): the more likely items are to be recognized or recalled via cues, the shorter is the response time in a linear relationship. This suggests that the free recall retrieval mechanism is the same in recognition and cued recall; and that the mechanism is a reactivation mechanism (Tarnow, 2008; Tarnow, 2009). The only difference is in the time scale: the largest reactivation time for recognition and cued recall varies between 0.2-1.8 seconds while the largest retrieval time in free recall is 7.2 seconds from extrapolating the line in Fig. 3.

**General Discussion**

This work implies that there are only two types of short term memory – working memory and reactivation.

I have previously suggested that the recognition and cued recall mechanism is one of reactivation (Tarnow, 2008), specifically, synaptic exocytosis (Tarnow, 2009). I arrived at this suggestion by showing that the response time increases linearly with the probability of correct recognition / cued recall, just like synaptic exocytosis. The probability of correct recognition / cued recall decays logarithmically with time, just like synaptic endocytosis. In the present contribution I showed that in free recall the average retrieval time increases linearly with the probability of correct recall and that the time scale for free recall retrieval is the same order as the time scale of recognition and cued recall. To show the second step, that the decay is logarithmic in time just like synaptic endocytosis, I would need a dataset from a delayed free recall experiment in which the delay was varied.

The consequences are important not only for the understanding of short term memory but also for an important area of clinical practice: the construction of short term memory tests. For example, if we want to test for the presence of Alzheimer's disease, the existing number of short term memory tests is huge (see, for example, http://www.nia.nih.gov/research/cognitive-instrument and select "memory" as the cognitive domain and about 88 tests can be found as of June 30, 2015). But if there are only two types of damage possible to short term memory – damage to working memory and damage to reactivation - this should cut down the number of tests to two: one for the presence of reactivation issues and another for the presence of working memory issues. Whether the tests measure recognition, cued recall or free recall should be irrelevant. One of the known facts about short term memory damage in Alzheimer's disease comes from the three word recall test of the MMSE (Folstein & Folstein, 1998): early Alzheimer's can be identified by delayed free recall of these words and later stage Alzheimer's can be identified by the inability to add the three words to working memory (Ashford et al, 1989). There may be no other short term memory test necessary beyond refinements of the MMSE.

Finally, let's return to the serial position curve in Fig. 1. If free recall, recognition and cued recall probe the same two types of memory, how come we never see anything so complex in recognition or cued recall? The answer is that in recognition and cued recall, working memory and reactivation are always probed together, they are never separated out.

In addition, since the recognition and cued recall probes also display items, those items keep displacing items in working memory largely removing the effects of working memory. In free recall, however, working memory is separate from reactivated memory – working memory is recalled in a first stage and only then is reactivated memory recalled (see Tarnow, 2015). The linear relationships presented here presumably apply only to the second stage.

## APPENDIX I: The recall data

The Murdock & Okada (1970) data can be downloaded from the Computational Memory Lab at University of Pennsylvania (http://memory.psych.upenn.edu/DataArchive). Lists of twenty words were read to groups of subjects to the accompaniment of a metronome (to fix the word presentation rate) after which the subjects wrote down as many of the words as they could remember in any order for a period of 1.5 minutes. There were two presentation rates: 1 or 2 words per second.

In Kahana et al (2002) two groups of 25 subjects, one older (74 mean age) and one younger (19 mean age), studied lists of ten words displayed on a computer screen. Immediately following the list presentation participants were given a 16-s arithmetic distractor task and only after this delay were the subjects asked to recall the words. This data can also be downloaded from the Computational Memory Lab at University of Pennsylvania (http://memory.psych.upenn.edu/DataArchive).

For both sets of data, trials with errors were discarded in order to keep the number of items to be remembered or having been recalled well defined. Data points with fewer than 10 measurements were not included in the graphs to minimize noise.